\documentclass[aps,prb,reprint,amssymb,longbibliography,preprintnumbers,twocolumn,amsmath]{revtex4-2}
\usepackage{graphicx}
\usepackage{amsmath,bm}
\usepackage{hyperref}
\usepackage{physics}
\usepackage[final]{showlabels}
\newcount\Comments  
\usepackage{color}
\definecolor{darkgreen}{rgb}{0,0.5,0}
\definecolor{purple}{rgb}{1,0,1}
\newcommand{\kibitz}[2]{\ifnum\Comments=1\textcolor{#1}{#2}\fi}

\begin{document}

\title{Edge-state and bulk-like laser-induced correlation effects in high-harmonic generation from a linear chain}

\author{Simon Vendelbo Bylling Jensen}
\affiliation{Department of Physics and Astronomy, Aarhus
University, DK-8000 Aarhus C, Denmark}

\author{Lars Bojer Madsen}
\affiliation{Department of Physics and Astronomy, Aarhus
University, DK-8000 Aarhus C, Denmark}

\date{\today}

\begin{abstract}
We explore the significance of laser-induced correlation effects for high-order harmonic generation (HHG) in a linear chain model of a generic band gap material for both single-IR-pulse and VUV-pump--IR-probe HHG schemes. We examine various pump-pulse excitations, and catagorize signatures of laser-induced correlation effects into two classes related to edge-state and bulk-like effects.  The relative importance of the edge-state effects decrease with increasing system-size, while the bulk-like effects remain. We identify regimes where these effects may alter the harmonic yield by an order of magnitude, but also regions where they may be neglected. We characterize the underlying laser-induced dynamics of the correlational mechanisms, using both an electronic population-description in a static field-free basis, and comparing to a bandstructure-picture with laser-induced time-dependent energy shifts. From this we provide a guideline on when to account for laser-induced electron correlations and give an estimate of the time-scale in which the laser-induced electron-electron interactions can cause correlational changes in the harmonic spectra. We find this timescale to be a few tens of femtoseconds.
\end{abstract}
\maketitle

\section{Introduction}
High-harmonic generation (HHG) has been utilized for decades to produce coherent ultraviolet light of ultrafast duration. The radiation inherits descriptive system-specific characteristics through the ultrafast electron dynamic of the HHG mechanism \cite{PhysRevA.66.023805,PhysRevLett.98.203007,Li1207,Baker424,PhysRevLett.94.053004,Itatani2004,Schubert2014,Luu2015,Garg2016,PhysRevLett.115.193603}. For HHG in a gaseous system, the mechanism was explained through the three-step model that can be most simply rationalized in terms of  electron trajectories in real-space involving ionization, propagation and recombination steps \cite{PhysRevLett.68.3535,PhysRevA.49.2117,PhysRevLett.71.1994}. For multielectron excitations in gases, an additional plateau arrives in the HHG signal as a characteristic of  laser-induced electronic correlations, which is described by nonsequential double-recombination \cite{PhysRevLett.98.043904,Hansen2016,PhysRevA.96.013401}. Recently HHG in solids has been under intense investigation. Compared to gaseous systems, a different scaling with laser intensity and a multiplateau structure of the HHG signal  is observed \cite{Ghimire2011}. The mechanisms of HHG in solids are subject to vivid discussion \cite{PhysRevA.85.043836,PhysRevA.91.013405,Hohenleutner2015,PhysRevA.94.063403,You2017,PhysRevLett.122.193901,PhysRevA.95.043416,PhysRevLett.113.213901,PhysRevLett.113.073901,PhysRevLett.116.016601,PhysRevLett.124.153204,PhysRevA.101.053411,Lakhotia2020}. Often a crystal-momentum-space description, which involves intraband dynamics and interband transitions, is applied.  This description shares elements of the three-step model such as excitation, propagation and recombination, but in crystal-momentum space \cite{PhysRevLett.113.073901}. The majority of physics models for HHG in semiconductors \cite{Ghimire2011,Schubert2014,Luu2015,Garg2016,PhysRevA.85.043836,PhysRevA.91.013405,Hohenleutner2015,PhysRevA.94.063403,You2017,PhysRevLett.122.193901,PhysRevA.95.043416,PhysRevLett.113.213901,PhysRevLett.113.073901} are based on the independent-electron approximation and describe the HHG mechanism as a result of electron population being transferred within or across bands of a static bandstructure. Such an independent-electron approximation was shown by \textit{ab initio} approaches to be accurate when describing \textit{single}-IR-pulse driven HHG in semiconductors and band gap (BG) model systems \cite{PhysRevLett.118.087403,Tancogne-Dejean2017,PhysRevA.96.053418,PhysRevA.102.033105}, whether being doped \cite{PhysRevA.99.013435}, finite \cite{PhysRevA.97.043424}, containing vacancies \cite{PhysRevResearch.2.013204}, disordered \cite{PhysRevLett.125.083901} or having a topological nature \cite{PhysRevLett.120.177401}. In addition to the investigations in these systems that are weakly correlated under field-free conditions, aspects of HHG in intrinsically strongly correlated  materials have also been explored. In this latter case in terms of effective-Hamiltonian models \cite{PhysRevLett.124.157404,PhysRevLett.121.057405,Silva2018,PhysRevB.99.184303,PhysRevLett.121.097402,PhysRevB.102.081121,PhysRevB.103.035110}.

It is intriguing that while the multiplateau structure of the HHG signal in the atomic case is an electron-correlation effect, the multiplateau HHG structure of the response of a {\it solid} appears as a result of the bandstructure, and is not altered significantly by laser-induced correlations in the case of a \textit{single} pulse HHG schemes. To this extent, it is interesting to see whether one can create laser-induced correlation effects in the HHG mechanism in weakly correlated nanostructured or bulk BG materials. 
Very recently it was shown \cite{jensen2020edgestateinduced} that significant laser-induced correlations can indeed be triggered in a linear chain model of a finite-size  semiconductor, when exposed to a VUV-pump--IR-probe scheme with a controlled time-delay between the pump and the probe pulse. By resonantly exciting the edge states (ESs) of the finite system, it was found that the ESs, which are localized in real-space, cause real-space density fluctuations that necessitate a correlated description. It was found that while the ES-induced correlation can give both significant enhancement and decrease in specific regions of the HHG spectrum for the finite-system, the relative importance of the localized ESs decrease as the system-size increases. Accordingly the associated laser-induced correlation effect is strongly suppressed in the bulk limit. 

In this work we go beyond the consideration of electron correlation induced by the laser-mediated population of ESs. We do so by varying the frequency of the pump pulse and we find laser-induced correlation effects, that persist in the bulk-size limit, which we therefore characterize with a crystal-momentum-space description.  When it comes to the possible roles of laser-induced correlation,  many interesting questions are still unexplored and we seek to address some of these in the following: How can the laser-induced correlation effects be included in a bandstructure picture and do different mechanisms arrive with different properties? To what extent can laser-induced correlation effects alter the HHG spectra for HHG by a single IR pulse or by a VUV-pump and an IR-probe pulse?  What time-scale is associated with the mechanisms of laser-induced correlations? Furthermore, we aim to provide a guideline for domains where one can neglect laser-induced correlation effects and accurately apply the independent-electron approximation for a generic semiconductor. In order to consider these questions we will first provide a short description of the laser-induced electron-correlation mechanisms in one-color HHG from a model of a BG material with more than one valence band (VB), where it is known from the literature that signatures of laser-induced correlation effects in the HHG spectra are at high order and typically out of range for experimental detection \cite{PhysRevB.102.174302,PhysRevA.96.053418, PhysRevLett.118.087403}.  Supplementing this discussion, we supply a method to reveal the underlying mechanics through correlation-induced electron-populations considered in the static field-free ground state bandstructure. Hereafter, we employ a pump-probe scheme for HHG using VUV and IR pulses with a variable time-delay between them. From these studies we identify two distinct types of laser-induced electron correlation effects, being ES or bulk-like effects. We distinguish the correlational dynamics through both a static bandstructure model with correlation-induced electronic populations, and comparing with an analysis of the laser-induced correlational energyshifts if considering a time-dependent basis. Utilizing the capability of time-resolution through pump-probe HHG spectroscopy we are able to characterize the time-scale of the laser-induced correlational dynamics. We thus provide new insight into the characteristics of ultrafast beyond-mean-field electron-electron interactions.

The paper is organized as follows. In Sec.~II, the theoretical model and methods are presented. In Sec.~III, the results are discussed. Section IV gives the conclusions. Atomic units are used throughout unless otherwise indicated.

\section{Theoretical Model and Methods}

In this work, time-dependent density functional theory (TDDFT) \cite{PhysRevLett.52.997} is applied within the local spin-density approximation.  It is self-consistent, accounts for electron-electron interactions and includes contributions from all bands in the bandstructure \cite{ComputationalStrongFieldQuantumDynamics}. We do not account for macroscopic light propagation effects such as dephasing and absorption, even though such effects may modify the HHG spectra \cite{PhysRevLett.125.083901}.  These effects can be suppressed through their dependence on target thickness, so the present  approach is valid for thin target materials, which are feasible to produce, see, e.g., Refs.  \cite{Liu2017,Garg2016,PhysRevLett.125.097004,PhysRevB.96.195420,Yoshikawa736}.

 We consider a model of a BG material with more than one filled VB to include the possibility of interband contributions from lower energy VBs. The lower VB electrons do often not contribute significantly to the HHG signal and are hence often neglected to obtain effective two-band or $N$-band models with one VB and one or more conduction bands (CBs). Since, single-VB and two-VB models, in some cases, exhibit different indications of laser-induced electron correlations~\cite{PhysRevB.102.174302}, we include all VBs in the present calculations.

The applied model exhibits a well-converged bulk-system response for increasing number of atoms, and has been widely used as a benchmark when examining HHG \cite{PhysRevA.97.043424,PhysRevA.96.053418,PhysRevA.99.013435,PhysRevResearch.2.013204,jensen2020edgestateinduced,PhysRevB.102.174302,PhysRevA.102.033105}. 
We consider a linear chain of $N$ ions, each with nuclear charge $Z = 4$. The ion positions $x_i$, separated by the lattice constant $a = 7$, are given as
$x_i = \left[ i - \frac{1}{2} \left( N - 1 \right)\right] a. $
The static ionic potential is
\begin{equation}
v_{\mathrm{ion}} \left( x\right) = - \sum_{i=0}^{N-1} \frac{Z}{\sqrt{\left(x-x_i \right)^2 + \epsilon} } \label{eq:vion}
\end{equation}
with softening parameter $\epsilon = 2.25$. The softening parameter is chosen to soften the Coulomb singularity. We aim to capture the physics of 3D electrons driven in the polarization direction of a linearly polarized electromagnetic field, rather than a true 1D system, all without unduly distorting electron-electron correlations, see, e.g., Ref.~\cite{PhysRevLett.68.2905}.  Calculations have been made with systems of varying length, from nanoscale systems to bulk systems. In the following, the  $N=80$ ($\sim 30$ nm) sample is denoted as a finite nanoscale system, as it matches obtainable diameters in production of, e.g., ZnO nanowires \cite{Grinblat2014}. Such nanostructures are of interest with the scope of utilizing enhancement from plasmons \cite{Cox2017}. For comparison the response of a $N=300$ ($\sim 111$ nm) sample will be denoted as the response of a bulk system as this system size is well above the size of $N=220$ where all previous single-pulse and pump-probe-pulse simulations have shown convergence to the bulk response \cite{PhysRevA.102.033105,jensen2020edgestateinduced}.

In the absence of the external time-dependent field, the many-body interacting fermionic system may be  represented by an auxiliary system of noninteracting Kohn-Sham (KS) orbitals through the Hohenberg-Kohn theorem~\cite{Hohenberg1964}. The KS orbitals $\varphi_{\sigma,i}$, with spin $\sigma = \lbrace \uparrow, \downarrow \rbrace$ satisfy the KS equation
\begin{equation}
 \left\lbrace -\frac{1}{2} \pdv[2]{x} + v_{\mathrm{KS}} \left[n_\sigma\right] \left(x\right)  \right\rbrace \varphi_{\sigma,i} \left(x \right)  = \varepsilon_{\sigma,i} \varphi_{\sigma,i} \left(x \right),  \label{eq:kseq}
 \end{equation}
with the static KS potential
\begin{equation}
v_{KS}\left[ \left\lbrace n_\sigma \right\rbrace \right] \left( x \right) = v_{\mathrm{ion}} \left(x\right) + v_H \left[ n \right] \left( x \right) + v_{xc}\left[ \left\lbrace n_\sigma \right\rbrace \right] \left( x \right). \label{eq:kspot}
\end{equation}
Here  the first term represents the interaction with the ionic lattice and is given by Eq.~\eqref{eq:vion}. The second term describes the Hartree potential and is given by
\begin{equation}
 v_H \left[ n \right] \left( x \right) = \int dx' \frac{n\left( x'\right)}{\sqrt{\left(x-x' \right)^2 + \epsilon}}.
\end{equation}
The last term in Eq.~\eqref{eq:kspot} describes the exchange-correlation potential and is given by the following expression in the local spin-density approximation 
\begin{equation}
v_{xc}\left[ \left\lbrace n_\sigma \right\rbrace \right] \left( x \right) \simeq v_{x}\left[ \left\lbrace n_\sigma \right\rbrace \right] \left( x \right) = - \left[ \frac{6}{\pi} n_\sigma (x) \right]^{1/3}.
\end{equation}
The spin density and the total density are given, respectively, as
 $n_\sigma \left(x\right) = \sum_{i=0}^{N_\sigma -1} \abs{\varphi_{\sigma,i} \left( x \right)}^2$ and  $n\left(x\right) = \sum_{\sigma = \uparrow, \downarrow} n_\sigma \left(x\right)$
with $N_\sigma$ being the number of electrons with spin $\sigma$. In this work we consider charge- and spin-neutral systems such that $N_{\uparrow, \downarrow} = Z N /2$. We have checked that the conclusions of the present work, including those related to field-induced dynamics,  remain unaffected by using another functional \cite{PhysRevLett.77.3865}. 

For propagation in the presence of external laser pulses, we apply the time-dependent KS equation (TDKSE)
\begin{align}
i \pdv{t} \varphi_{\sigma,i} \left(x,t \right) =& \bigg\lbrace  - \frac{1}{2}\pdv[2]{x} - i A\left(t \right) \pdv{x} \nonumber \\
& \ \ +  \tilde{v}_{\mathrm{KS}} \left[n_\sigma\right] \left(x,t \right) \bigg\rbrace \varphi_{\sigma,i} \left(x ,t \right) \label{eq:kseqtime}
\end{align}
using  the vector potential $A\left(t\right)$ to describe the electromagnetic field and the time-dependent KS potential
\begin{equation}
\tilde{v}_{KS}\left[ \left\lbrace n_\sigma \right\rbrace \right] \left( x ,t\right) = v_{\mathrm{ion}} \left(x\right) + v_H \left[ n \right] \left( x ,t \right) + v_{xc}\left[ \left\lbrace n_\sigma \right\rbrace \right] \left( x,t \right) \label{eq:kspottime},
\end{equation}
in which the Hartree and exchange-correlation potentials inherit the time-dependence of the electron density. A complex absorbing potential was added to prevent electrons that reach the boundary of the simulation box to backscatter and contribute to unphysical interaction, for this method see, e.g., Ref.~\cite{KOSLOFF1986363}.  We apply the Crank-Nicolson method for time-propagation with a predictor-corrector step ~\cite{ComputationalStrongFieldQuantumDynamics} and work in the adiabatic approximation for the TDKSE~\cite{ullrich2012time}.  First, the ground state is found by imaginary time propagation of Eq.~\eqref{eq:kseqtime} renormalizing the orbitals at each time step. For imaginary time-propagation, the system is propagated until convergence with timestep $\Delta t = 0.5$ a.u., whereas a timestep of $\Delta t = 0.1$ a.u. is applied for the real time propagation, until well beyond the end of the driving pulse. The converged results are found on a spatial grid with spacing $\Delta x = 0.1$ a.u. The gridsize applied for convergence, is of $17000$ and $67500$ grid points for the nanoscale and bulk system, respectively. Here the $560$ a.u. ($2100$ a.u.) nanoscale (bulk) system is located at the center of the simulation box and occupy less than a third of the simulated box.

\begin{figure}
\includegraphics[width=0.49\textwidth,height=8cm]{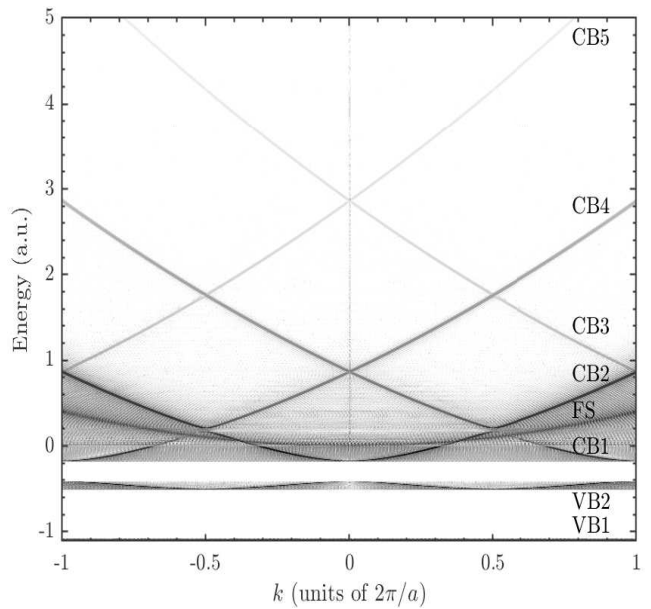}
\caption{Part of the bandstructure resulting from plotting the norm squared of the Fourier transformed occupied and unoccupied KS-orbitals for the finite-size system ($N=80$), as a function of their corresponding energies and the Fourier variable $k$. The occupied valence bands are labelled as VBi with i $=1,2$ and the unoccupied conduction bands are labelled as CBi with i $=1,2,3,4,5...$. Due to the surrounding vacuum in the simulation box, the free-space (FS) dispersion of $k^2/2$ is also visible.} \label{fig:1}
\end{figure}

With the obtained ground state, the KS-potential is evaluated through Eq.~\eqref{eq:kspot}, and the Hamiltonian in Eq.~\eqref{eq:kseq} is diagonalized. A bandstucture can hereafter be constructed by the norm square of the Fourier transformed KS orbitals. Such bandstructures are available for the considered model system in the literature in, e.g., Refs. \cite{PhysRevA.97.043424,PhysRevA.96.053418,PhysRevA.99.013435,PhysRevA.102.033105,PhysRevResearch.2.013204,jensen2020edgestateinduced,PhysRevB.102.174302}. 
As seen from Fig.~\ref{fig:1}, the bandstructure consists of two VBs, denoted VB1 and VB2, which are fully occupied by electrons and separated by $0.570$  ($\sim 15.5$ eV). The system has a BG energy of $0.239$ ($\sim 6.5$ eV), whereafter there are multiple unpopulated CBs. As the system is of finite size, it contains a few ESs which can typically be found with energies just below each VB and CB. Finally, one can see a free-space (FS) parabola, which signifies that the system is finite and that electrons can escape from the finite linear chain into vacuum. In previous work, it was shown that these FS electrons do not contribute significantly to the HHG yield~\cite{PhysRevA.97.043424}.

Using the ground state as the initial state, we apply a driving (subscript $d$) IR pulse described by  $A_{d}\left(t \right)$. In the pump-probe simulations, we add a pump preexcitation (subscript $p$) VUV pulse $A_{p}\left(t \right)$. In general the linearly polarized pulses in the dipole approximation are described by the vector potential
\begin{equation}
A\left(t \right) = A_{p}\left(t \right) + A_{d}\left(t \right).  \label{eq:emfield}
\end{equation}
Their explicit forms are
\begin{align}
A_p\left(t \right) &= A_{p} \sin^2\left( \frac{\omega_p t}{2 n_p}\right) \sin \left( \omega_p t - \phi_p \right), \ \ \ 0 < t < \frac{2\pi n_p}{\omega_p} \nonumber \\
A_d\left(t \right) &= A_d \sin^2\left( \frac{\omega_d \left(t-\tau \right)}{2 n_d}\right) \sin \left( \omega_d \left(t-\tau \right) - \phi_d  \right),  \nonumber \\  &\ \tau + \pi \left(\frac{n_p}{\omega_p} - \frac{n_d}{\omega_d} \right) < t < \tau + \pi \left(\frac{n_p}{\omega_p} + \frac{n_d}{\omega_d} \right). \nonumber 
\end{align} 
Here $\omega_i$ is the angular frequency, $n_i$ is the number of cycles  and $\phi_i$ is the carrier envelope phase of the pulse $i =  \lbrace d,p \rbrace$. Furthermore, $\tau$ is the peak-to-peak time delay between the two pulses. Throughout this work, we apply a $n_d=15$ cycle driving pulse with a frequency of $\omega_d = 0.023$ ($\lambda_d \simeq 2 \mathrm{ \mu m}$) with $A_d = 0.24$ or $F_d = 0.00552$ corresponding to an intensity of $\simeq 10^{12} \mathrm{W/cm^2}$. The parameters of the preexcitation pump pulse will be varied and the effect of this variation is examined in Sec.~\ref{Sec:two color}.

As we aim to compare uncorrelated and correlated electron dynamics in this work, we rely on the wording used in the literature, see, e.g., Refs.~\cite{PhysRevLett.118.087403, PhysRevLett.121.097402,Tancogne-Dejeaneaao5207}. When discussing uncorrelated, independent-electron dynamics, we refer to the approximation $\tilde{v}_{KS}\left[ \left\lbrace n_\sigma \right\rbrace \right] \left( x ,t\right) \simeq \tilde{v}_{KS}\left[ \left\lbrace n_\sigma \right\rbrace \right] \left( x ,0 \right) =v_{KS}\left[ \left\lbrace n_\sigma \right\rbrace \right] \left( x\right)$. This corresponds to uncorrelated, independent-electron dynamics in the sense that it only includes the electron-electron interaction from electrons in the initial state, which is the electronic ground state. This ground-state electron-electron interaction is evaluated to form a time-independent effective potential, wherein all electrons move independently.
We refer to the dynamics being correlated if we go beyond this approach and allow for a dynamical electron-electron interaction throughout the calculation, retaining the time-dependence in the Hartree and exchange-correlation potential as described by Eq.~\eqref{eq:kspottime}. Thus we can express the Hamiltonian that accounts for correlation during the dynamics, ${H}_c$, from the TDKSE, in terms of the  Hamilton of the uncorrelated system, ${H}_u$, as  
\begin{equation}
{H}_c = {H}_u + \delta v_{KS} \left[\lbrace n_\sigma \rbrace \right] \left( x ,t \right), \label{eq:Hamilton_cor}
\end{equation}
with the laser-induced correlation term
\begin{equation}
\delta v_{KS} \left[\lbrace n_\sigma \rbrace \right] \left( x ,t \right) = \tilde{v}_{KS}\left[ \left\lbrace n_\sigma \right\rbrace  \right] \left( x ,t\right) - \tilde{v}_{KS}\left[ \left\lbrace n_\sigma \right\rbrace  \right] \left( x ,0\right), \label{eq:cor}
\end{equation}
which describes how the laser-induced changes in the density lead to a time-dependent change in the interaction between the electrons. 

The HHG spectra can be found as 
\begin{equation}
 S(\omega) \propto \abs{\int dt J\left( t \right) e^{-i\omega t}}^2,
 \end{equation}
 with the time-dependent current
\begin{equation}
J(t)= \sum_{i,\sigma} \int dx \Re \left[\varphi^\ast _{i,\sigma} \left( x,t \right) \left( -i \pdv{x} + A\left(t\right)\right) \varphi_{i,\sigma} \left( x,t \right)  \right].
\end{equation}
To compare the spectra for single-pulse and pump-probe, two-pulse calculations on an equal footing, we apply the same window function, when evaluating the Fourier transforms. The window function attains the value $1$ before the center of the driving pulse, whereafter it follows a $\cos^8$ decay, inspired from similar pump-probe settings of Ref.~\cite{PhysRevA.98.053401}.

\section{Results and Discussion}
For completeness, we start in Sec.~III.A with a summary regarding laser-induced electron correlation effects in single-pulse HHG from BG materials. We supplement the well-known conclusions by applying a method to track the electron population in individual VBs and CBs. In Sec.~III.B, we then introduce a  preexitation pump pulse to examine laser-induced correlation effects in HHG in a pump-probe scheme. The results following from a scan of the parameters of the preexcitation pump pulse allows us to identify two distinct types of laser-induced correlation effects, finite-size edge-state effects and and bulk-like effects, with accompanying analysis of the HHG spectra for different preexcitation pump frequencies. To analyse the dynamics, the electron population was tracked along with any laser-induced changes of the bandstructure. For a temporal analysis the delay between the two pulses was varied to see signatures of laser-induced correlations in the HHG spectra.

\subsection{Laser-induced correlations effects in HHG with a single driving pulse \label{Sec:one color}}
\begin{figure}
\includegraphics[width=0.49\textwidth]{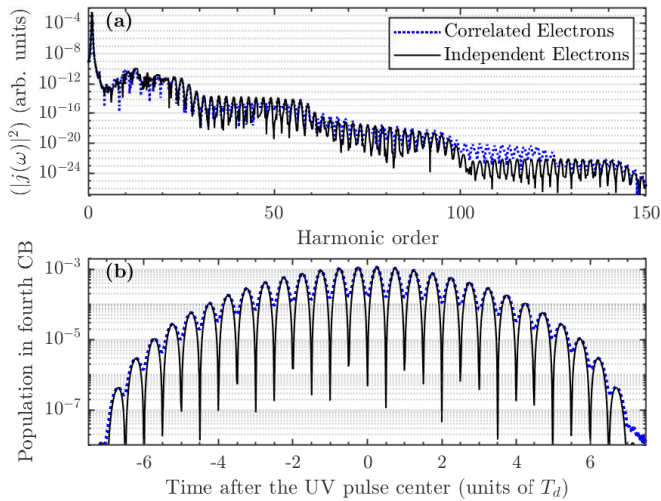}
\caption{(a) HHG spectra for a finite-size system ($N=80$) when applying an $n_d = 15$-cycle IR driving pulse with $\omega_d = 0.023$ and $F_{0,d} = 0.00552$ with independent-electrons and correlated electrons. (b) Electron population in CB4 as a function of time in units of the driving field duration $T_d = 2 \pi / \omega_d$.} \label{fig:2}
\end{figure}

For a generic semiconductor with more than one filled VB for HHG, laser-induced correlation effects were shown to be modest and out of range for experimental detection when applying only a single driving IR pulse~\cite{PhysRevB.102.174302,PhysRevA.96.053418, PhysRevLett.118.087403}. For descriptions of laser-induced correlations in single-VB semiconductors or metals see Ref.~\cite{PhysRevB.102.174302}. In our model, as previously seen in Refs.~\cite{PhysRevA.96.053418,PhysRevB.102.174302}, laser-induced electron correlation effects appear as an enhancement of harmonics in the high frequency region, corresponding to transitions from the fourth CB (CB4) to VB2 [see Fig.~\ref{fig:1}]. This enhancement can be seen in the HHG spectra of Fig.~\ref{fig:2}(a) starting at harmonic order $\sim 100$. The intensity of the signal at such high frequencies is generally too weak to be observed experimentally. Furthermore, the intensity at this frequency range is also on the verge of numerical noise, when simulated using a gradient based functional \cite{PhysRevLett.77.3865} for TDDFT. We have found such a laser-induced correlational enhancement of the fourth plateau for all system sizes above the limit of atomic behavior ($N \geq 60$; see Ref.~\cite{PhysRevA.97.043424}).  

\subsubsection{Hauling-up effect}

A mechanism to explain such laser-induced electron-correlation enhancement of higher HHG plateaus is the "hauling-up" effect \cite{PhysRevA.98.023415}. The hauling-up effect arrives from additional correlation-induced couplings to higher CBs. The emergence of additional coupling terms can, e.g.,  be seen explicitly when expanding the orbitals in Houston states $\ket{\tilde{\phi}_{nk_0}}$ for band $n$ and crystal momentum $k_0$, see, e.g., Ref.~\cite{PhysRevB.33.5494}. In the TDDFT sense, this corresponds to expanding the KS-orbitals, in the instantaneous eigenstates of the independent-electron KS-equation. In doing so, the complex expansion coefficients  $\alpha_{bk_0}^m(t)$ of the Houston-like eigenstates will be time-dependent and determined from
\begin{align}
\label{alpha}
i \pdv{t} \alpha_{bk_0}^m \left(t\right) =& \sum_n \alpha_{bk_0}^n \left(t\right) e^{i \int \varepsilon_{mn}\left[k\left(t'\right)\right] dt'}  \bigg( d^{mn}_{k(t)} E(t) \nonumber\\
& + \bra{\tilde{\phi}_{mk_0}} \delta v_{KS} \left[\lbrace n_\sigma \rbrace \right] \left( x ,t \right) \ket{\tilde{\phi}_{nk_0}} \bigg)
\end{align}
with energy differences $\varepsilon_{mn}(k(t))$, and transition dipole elements $d^{mn}_{k(t)}$ between bands of indexes $m$ and $n$ at $k\left(t\right) = k_0 + A\left(t\right)$. The second term in Eq.~\eqref{alpha} contains the effect of laser-induced correlations as seen earlier in Eqs. \eqref{eq:Hamilton_cor}-\eqref{eq:cor}. In Ref.~\cite{PhysRevA.98.023415},  using a time-dependent Hartree-Fock approach, a term similar to the last one in Eq.~\eqref{alpha} was found to introduce additional couplings for VB electrons when transitioning to higher-lying CBs. Due to this, the formation of plateaus was reported at lower field strengths as a direct consequence of laser-induced electron correlations. Hence, the hauling-up effect provides a mechanism for alterations of the HHG spectra by increased population of electrons in the higher CBs.

\subsubsection{Population analysis}

To supplement the interpretation of these laser-induced electron-correlation mechanisms, we allude to the result in Fig.~\ref{fig:2}(b). Here the electron population of CB4  [see Fig.~\ref{fig:1}] is plotted as a function of time during the driving pulse.  The population is determined with a procedure similar to that used in Ref.~\cite{Tancogne-Dejean2017}, where at each timestep the orbitals are projected onto each KS orbital  of the initial ground-state (GS) as
\begin{equation}
P_{\sigma,j}(t) = \sum_{i} \sum_{\sigma'} \abs{\bra{\varphi_{\sigma',i}(t)} \ket{\varphi_{\sigma,j}^{\mathrm{GS}}}} \label{eq:proj}.
\end{equation} 
Now the projection of the k'th CB, denoted as CBk can be found by summation
\begin{equation}
N_{\mathrm{CBk}}(t)=\sum_{j\in \mathrm{CBk}} \sum_\sigma P_{\sigma,j}(t). \label{eq:proj_sum}
\end{equation} 
By comparing the electron population in all of the individual CBs (not shown) for the independent-electron and correlated approaches, we only observe a correlation-induced increase in population in the fourth and higher CBs. For CB4, populations obtained with the two approaches are depicted in Fig.~\ref{fig:2}(b), where the overall amount of excited electrons is higher for the correlated than for the independent-electron approach throughout the duration of the driving pulse, which is consistent with the anticipated effect of the hauling-up mechanism. To directly relate a higher population in CBk k$\ge 4$ to an increase in the associated part of the HHG signal is not trivial. The emittance of HHG depends on the quantum interference between recombinations or imperfect recombinations at multiple $k$-points~\cite{PhysRevLett.124.153204}. The increase in population in a high-lying CB is thus not directly related to an increase in HHG-signal. We do, however, observe a higher degree of excitation to the highest CBs, consistent with the correlational hauling-up mechanism, and comparing with the HHG spectra we similarly find indications of laser-induced correlations at the associated recombination energies, i.e., at orders higher than $\sim 100$ in Fig.~\ref{fig:2} (a).

In this analysis of the electron population, we project onto a basis that arrives from considering only the electron-electron interactions in the electronic ground state. As the system is excited by the laser pulse, it exhibits density fluctuations resulting in a time-dependent electron-electron interaction, which goes beyond the static electron-electron interaction from the electronic ground state; see Eqs.~\eqref{eq:Hamilton_cor}-\eqref{eq:cor}. Since the basis does not contain these new laser-induced correlations, it is important to check the basis for completeness. This can be done by adding up all of the projected populations, and assuring that any loss in population correspond to the amount of population which is ionized and absorbed by the surrounding absorbing potential. In such an analysis of a correlated system using an uncorrelated basis, it seems rational that since the laser-induced correlations are not included in the basis, they will appear in additional non-trivial time-dependent couplings between the states of the uncorrelated basis, as we saw in the hauling-up effect of Eq.~\eqref{alpha}.

\subsubsection{Laser-induced BG-energy reduction}

An interesting question is now, whether one, instead of interpreting laser-induced correlation effects in terms of couplings in the uncorrelated basis, could gain insights by simply analyzing the nature of the time-dependent basis that emerges upon accounting for laser-induced correlations - we refer to this latter basis as the correlated basis. If, e.g., the BG-energy would be reduced due to laser-induced correlations, this could have large consequences for the non-linear HHG process. An effective reduction of the BG is exactly what is expected when applying an intense laser pulse to a semiconductor in a single-electron Floquet-approach, see, e.g., Refs.~\cite{PhysRevB.12.1132,MIRANDA1983783}. However,  along with such a laser-dressed BG-reduction, an intense laser pulse will also cause excitations resulting in laser-induced time-dependent correlation effects beyond such single-electron picture. The behavior of the time-dependent correlated basis is expected to be affected by the non-trivial interplay of such effects. If the electron-electron interactions (beyond those of the static ground state) were to induce an energy shift through the KS-potential, it would require a significant change in the electron density. However, most of the density has been shown to remain in the ground state in multiple single-pulse HHG studies of different systems  \cite{PhysRevA.96.053418,PhysRevA.97.043424,PhysRevA.102.033105,PhysRevResearch.2.013204,PhysRevA.99.013435}. Since only a fraction of the orbitals of the highest VB change, the total density is not altered significantly, alluding to such correlational-induced BG-fluctuations being small. 

To determine the size of such a laser-induced BG-energy reduction we consider the time-dependent electron distribution at the instant corresponding to the center of the driving IR pulse. This electron density is applied to diagonalize the Hamiltonian in order to find a new bandstructure, which consist of the instantaneous eigenstates of the correlated electron KS-equation - the correlated basis. In the single pulse studies, we find the correlated bandstructure to be almost identical to the ground state bandstructure. Although, due to the slight change of the dynamic KS-potential, the BG energy will change from $10.62\  \omega_d$ to $10.35\  \omega_d$. The magnitude of the reduction is not as remarkable in this multielectron model as expected from the independent-electron model of Refs.~\cite{PhysRevB.12.1132,MIRANDA1983783}. The present change of the BG energy is not expected to significantly alter the HHG spectra. There could, nevertheless,  exist a range of laser parameters, where the HHG signal might be more responsive to laser-induced correlations if, e.g., the BG-fluctuation could lower the number of photons needed for the multiphoton excitation by the driving IR pulse. 

If we were to consider our system in a simpler model by neglecting, e.g., VB1, then this removal of VB1 would result in a removal of a large portion of the non-excited electrons since the VB1 electrons are not being significantly excited by the driving pulse. Neglecting the VB1 electrons  would hence increase the overall fraction of excited orbitals, and result in larger relative fluctuations of the density with the capability to affect the BG-energy more severely. We thus allude to caution when simplifying a multi-VB bandstructure with single-VB models, as single VB-systems have been reported to exhibit more correlation effects \cite{PhysRevB.102.174302}. 

 To investigate ways to enhance effects of laser-induced electron correlations in a controlled manner a possible way is by applying pump-probe schemes as will be explored in the next section.

\subsection{Laser-induced correlations effects in HHG in a pump-probe VUV-preexcitation -- IR-driving pulse scheme \label{Sec:two color}}
The aim of pump-probe HHG in solids is often to carefully select laser pulses with different frequencies, durations and delay as a mean of controlling the amount of excited electrons and promoting specific electronic pathways through the bandstructure for an increase in harmonic efficiency, see, e.g., the recent works in Refs.\cite{PhysRevB.100.214312,PhysRevA.101.033410}. When applying two pulses in a pump-probe manner, the experimental study of Ref.~\cite{Wang2017} shows that when a crystal is photoexcited and later driven, then the harmonic yield is reduced for bulk ZnO. However, laser-induced correlations significantly alter the HHG mechanisms if the frequency of the preexcitation pump pulse is suitably chosen as shown in the finite-size nanoscale regime in  Ref.~\cite{jensen2020edgestateinduced}. In that latter work, correlational enhancement is reported at a wide range of HH-frequencies entering the regime of the first plateau, where it would be feasible to detect experimentally. For the sake of experimental detection, we narrow the search of correlation effects to the first plateaus of the HHG spectra. In Sec.~II.B.1 we aim to give guidelines on how to evoke different kinds of laser-induced correlation effects. In order to obtain an intuition into how such laser-induced correlation effects could arrive, one can consider the correlated time-evolution operator $U_c$ compared to the uncorrelated $U_u$ of the Hamiltonians $H_c$ and $H_u$ of Eq.~\eqref{eq:Hamilton_cor}, which relate as
\begin{equation}
U_c (t+\Delta t, t) \simeq e^{-i \delta v_{KS} \left[\lbrace n_\sigma \rbrace \right] \left( x ,t \right) \Delta t} U_u(t+\Delta t, t). \label{eq:timediff}
\end{equation}
We see here that the laser-induced correlation supplies an additional phase, which through time will accumulate. Thus we expect to find an increase in laser-induced correlations when probing the system for a longer duration with, e.g., temporally separated pulses. If we consider, that the potential $\delta v_{KS} \left[\lbrace n_\sigma \rbrace \right] \left( x ,t \right)$ supplies a phase with the same magnitude as the BG-energy shift of the previous section, then a crude approximation would be that laser-induced correlations arrive after $\Delta E \Delta t \sim 2 \pi$ which would be at times larger than $\Delta t \sim 1000$ a.u. ($\sim 24$ fs). As we will observe later, this rough estimate fits rather well with the timescale of the laser-induced correlations.

\subsubsection{Bandstructure and transitions \label{Sec:two_a}}
\begin{figure}
\includegraphics[width=0.49\textwidth]{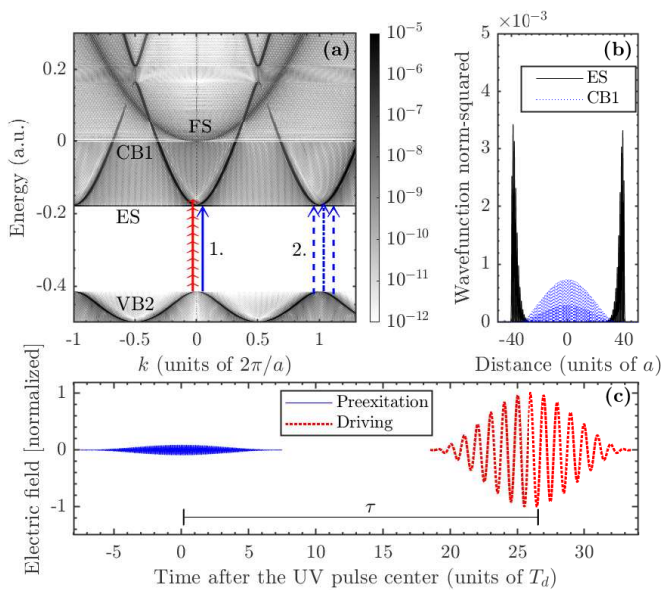}
\caption{(a) Zoom-in of VB2 and CB1 from Fig. \ref{fig:1}. The ESs are indicated below the CB by the horizontal line at $\approx - 0.18$. The free-space (FS) dispersion is visible. Arrows denote central photon energies, the smallest red arrows depicting the $11 \omega_d$ multiphoton transition, the solid blue arrow 1., illustrating the preexcitation pulse when resonant with the ES, and, arrows 2. denoting preexcitation pulses with  frequencies corresponding to energies close the BG frequency which allows for the blue dash-dotted transition to the lowest energy state in the CB or the blue dashed transition from the VB to the ES at another $k$. (b) Real-space norm-squared of the ES KS orbital  below the first CB and of the lowest energy state within the first CB for an $N=80$ system. (c) Pulse sequence with the VUV preexcitation and the IR driving laser pulse in units of the period of the IR laser, $T_d$, and with each pulse normalized to the peak field strength of the IR pulse. The pulses drive the transitions in (a), see text for parameters. 
} \label{fig:3}
\end{figure}

A part of the bandstructure is given in Fig.~\ref{fig:3}(a) depicting the occupied VB2 and unoccupied CB1 for the finite system. As previously reported in Refs.~\cite{jensen2020edgestateinduced,PhysRevA.102.033105} and magnified in Fig \ref{fig:3}, such finite systems contain a few ESs just below each VB and CB. The lowest energy ES below CB1 and the lowest energy state within CB1 can be compared in Fig.~\ref{fig:3}(b) for the $N=80$ system. For these lowest energy ES (CB1-state) we observe a delocalized (localized) orbital in momentum space and a localized (delocalized) norm-squared wavefunction in real space if comparing Figs.~\ref{fig:3}(a) and (b), respectively. The electromagnetic field from the definition in Eq.~\eqref{eq:emfield} is depicted in Fig.~\ref{fig:3}(c). Here a $n_p = 153$-cycle VUV preexcitation pulse is applied within the perturbative response regime with field strength $F_{0,p} = 0.0005$ ($I_d \simeq 9 \times 10^{9}$ W/cm$^2$). This pump pulse is followed by the driving pulse used in the single-pulse studies discussed above. The $11 \ \omega_d$-photon excitations for interband processes are depicted with red arrows in Fig.~\ref{fig:3}(a). Two interesting VUV preexcitation frequencies will be examined in detail. We will refer to these as the ES-preexcitation frequency with $\omega_{p,ES} = 0.235$ and the BG-preexcitation frequency with $\omega_{p,BG} = 0.239$. These correspond to resonant excitation from $k=0$ of VB2 to the two states of Fig.~\ref{fig:3}(b). The first of which, the ES-preexcitation, is depicted with the transition $1.$ in Fig.~\ref{fig:3}(a), and is a resonant coupling from the highest energy state within VB2 at $k=0$ to the lowest energy ES. The second, the BG-preexcitation, is tuned to the BG energy to drive multiple transitions, as depicted with $2.$ in Fig.~\ref{fig:3}(a). It provides a resonant coupling between the highest energy state in VB2 and the lowest energy state of CB1 at $k=0$, shown with the dashed-dotted arrow. Due to the ES being delocalized in momentum space, the latter frequency also couples states in VB2 to the ESs at a finite $k$, as shown with the dashed arrows. As in the recent study of Ref.~\cite{PhysRevB.100.214312}, we set the number of cycles of the VUV preexcitation pulse to $n_p = 153$. In this way, it has a pulse duration similar to that of the 15-cycle IR driving pulse. Due to the many cycles of the preexcitation pulses, their spectral width is rather narrow with a FWHM of the electric field of the pulse being less than $0.0025$ a.u., and thus the transitions induced by the VUV pump, as depicted with $1.$ and $2.$ in Fig. \ref{fig:3} (a) are well defined in energy space.  
To clearly identify the impact of the VUV preexcitation pulse on the HHG spectra, we will use the single-IR-pulse spectra as examined in Fig.~\ref{fig:2} as reference.

\subsubsection{Laser-induced correlation effects in HHG in temporally well-separated preexcitation VUV-pump and driving IR-probe pulses}
\begin{figure}
\includegraphics[width=0.49\textwidth]{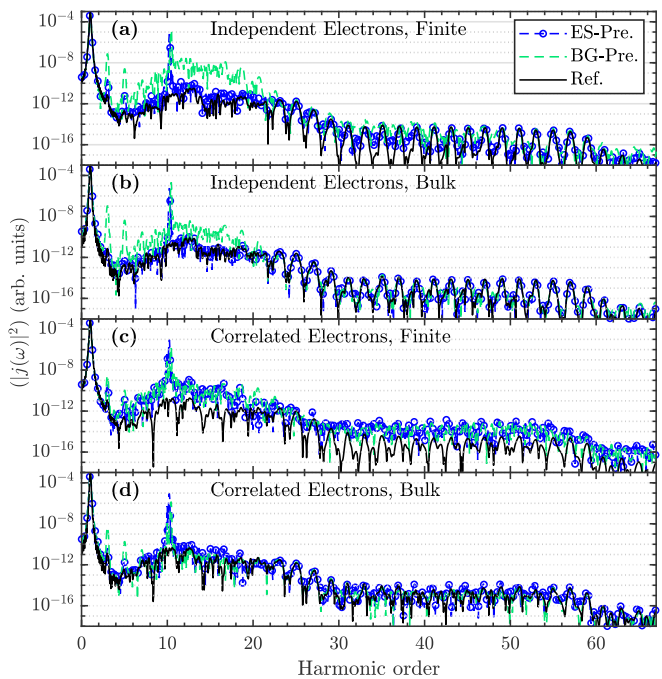} 
\caption{HHG spectra (a) for a finite-size system ($N=80$) with independent-electrons,  (b) for a bulk system with independent-electrons, (c) for a finite-size system including laser-induced correlations and (d) for a bulk system  including laser-induced correlations. The spectra are generated applying an IR-field ($n_d = 15$-cycle IR-pulse with $\omega_d = 0.023$ and $F_{0,d} = 0.00552$) to a system that has been preexcited with an ES preexcitation frequency (ES-Pre.), a system preexcited with a BG preexcitation  frequency (BG-Pre.) and a reference system without preexcitation (Ref.). 
The preexcited systems were prepared at  a time $\tau = 26 T_d$ earlier than the center of the IR pulse using an $n_p = 153$-cycle VUV-pulse with $\omega_{P,ES} = 0.235$ or  $\omega_{P,BG} = 0.239$, $F_{0,p} =0.0005$, and resonant with the ES or BG [see Fig.~\ref{fig:3}(a) and (b)]. The signals have been scaled by $N^{-2}$ to compare the systems on an equal footing.
} \label{fig:4}
\end{figure}
We first examine the case with temporally well-separated preexcitation VUV pump and driving IR probe pulses. As in the schemes of Refs.~\cite{PhysRevB.100.214312,jensen2020edgestateinduced} we apply a peak-to-peak delay of  $\tau = 26 T_d$, which is shorter than the spontaneous decay time of the carriers. Here $T_d = 2\pi/\omega_d$ ($\sim 6.6$ fs) is the period of the driving IR frequency. A  scan over VUV preexcitation frequencies has been made and we illustrate certain aspects by focusing on the two ES- and BG-frequencies corresponding to $1.$ and $2.$ of Fig.~\ref{fig:3}(a). The results of such calculations are depicted in Fig.~\ref{fig:4}. We note in passing that although not shown in the frequency range considered in the figure,  the hauling-up effect is present at the expected high frequencies in the correlated approaches [see Sec.~III.A.1]. We start by focussing on the case of the ES-preexcitation, as its effects need to be known and singled out to be able to perform an analysis of the results for the BG-frequency case as the latter frequency also populates resonantly to the ES from states lying lower in VB2  [see, e.g., Fig.~\ref{fig:3} 2. dashed lines]. When applying the ES-preexcitation, the states in CB1 can not be reached resonantly from the top of VB2 and the effects associated with ESs are singled out. We remind the reader of some of the results previously discussed in detail in Ref.~\cite{jensen2020edgestateinduced}: We see that a significant enhancement arrives from laser-induced correlations at a wide range of high harmonic frequencies for the finite-size system of Fig.~\ref{fig:4}(c). This enhancement is clearly an effect dominated by laser-induced correlations as it does not appear in Figs.~\ref{fig:4}(a) or (b) and it depends on system size, as it vanishes in the bulk limit of Fig.~\ref{fig:4}(d). The responsible mechanism is believed to be density-fluctuations which are initialized by resonantly populating the spatially localized ES. The relative importance of this ES-induced dynamics decreases with increasing system size where the ESs become of less significance when compared to the increasing number of bulk states. This reduces the relative density fluctuation and provides less dynamic laser-induced changes to the electron-electron correlation. The description of real-space density-fluctuations arriving from localized atom-like orbitals in a finite-size region, resembles methods and regimes of atomic and molecular systems. Even though only a few ESs are located in a narrow energy range in Fig.~\ref{fig:3}(a), the finite-size laser-induced correlational enhancement attributed to them is found at a wide range of VUV preexcittaion frequencies from $0.23 $a.u.$\leq \omega \leq 0.43$a.u. This insensitivity to the exact value of the preexcitation frequency of the VUV is due to the fact that the ESs are delocalized in momentum space and are available for resonant excitation from a wide range of VB2 states with different energies, and the fact that the energies of the ESs are highly dependent on the instantaneous electron-distribution of the sample. The ES energies thereby vary as a function of time when both preexciting and driving the system as well as in-between the two pulses.
 
When the VUV BG-frequency is applied, we observe quite remarkable effects of laser-induced correlations. Considering the results of the uncorrelated independent-electron calculations shown in  Figs.~\ref{fig:4}(a) and (b), a large enhancement of the HHG yield appears both below the BG energy (at $\sim 11 \omega_d$) and within the first plateau (at $\simeq 11-21 \omega_d$), consistent in both the finite and bulk system. However, these effects are severely reduced by laser-induced correlations as seen by the results in Figs.~\ref{fig:4}(c) and (d). The remaining finite-size enhancement in Fig.~\ref{fig:4}(c) is attributed to the transitions to the ES, that are still available for resonant excitation through the $2.$ dashed transition of Fig.~\ref{fig:3}(a). Thus laser-induced correlations seem to provide a large reduction in the HHG spectra when applying the BG-preexcitation. If one did not include laser-induced correlation, one would overestimate the HHG signal. We have checked that the magnitude of this overestimation corresponds to overestimating the laser intensity by more than an order of magnitude. As this over-estimation of the yield in the independent-electron approach is independent of system size, we seek to describe it as a bulk-like effect through an analysis in momentum space. A crucial point to make from this is that typical theoretical methodologies used for the description of HHG in solids, such as the semiconductor Bloch equations (SBE) implemented in their simplest form, rely on the independent-electron approximation. The results in Fig.~\ref{fig:4} show that this approximation can break down in pump-probe settings as it may overestimate the signal by an order of magnitude. 

\subsubsection{Laser-induced energy shifts with  VUV preexcitation }
Comparing the spectra for the BG-preexcitation and the ES-preexcitation in Fig.~\ref{fig:4}, we see that in the independent-electron approximation they provide widely different spectra. However, in the correlated case the spectra become almost identical. A reason for this can be found, if we are reminded of the discussion of Sec.~\ref{Sec:two_a}. Here the spectrum of states, which is populated by the different preexcitations is different if the bandstructure is static, as in the independent-electron approximation. However, as the two preexcitation frequencies, $\omega_{p,ES}$ and $\omega_{p,BG}$ are close in energy, a fluctuation of the BG energy, allowed in a dynamical calculation, might make both preexcitation frequencies span a similar range of transitions. To see whether this is the case, we take the density of the finite system at the center of each preexcitation pulse, construct the KS potential and diagonalize the system. We find that the BG of $0.2393$ a.u. is reducing to $0.2294$ a.u. during the ES-preexcitation with $\omega_{p,ES}$, and reducing to $0.2341$ a.u. during the BG-preexcitation with $\omega_{p,BG}$. These shifts in energy are larger than the spectral width of ($< 0.0025$ a.u.) for the preexcitation pulses. It is thus clear that a reason for the similar HHG spectra when including laser-induced correlation as seen in Figs. \ref{fig:4}(c) and (d) is that both pulses probe similar parts of the dynamical bandstructure and initialize a similar shift of the BG-energy.
Such BG-energy shifts cause the analysis based on a static bandstructure to be challenged as both preexcitation pulses are able to populate a wider spectrum of states than one might expect from a static bandstructure picture. If we consider the evolution of the BG energy during the driving pulse instead, it fluctuates for both cases in range between $0.222$ a.u. and $0.237$ a.u., with a mean of $0.2334$ a.u. for the ES-preexcitation and a mean of $0.2325$ a.u. for the BG-preexcitation. As a result of the BG-energy shifts during the driving pulse, the BG energy will at certain times reduce to below $10 \omega_d$, making a new excitation-channel possible for the driving pulse. 

\subsubsection{Population analysis with VUV preexcitation}
\begin{figure}
\includegraphics[width=0.49\textwidth]{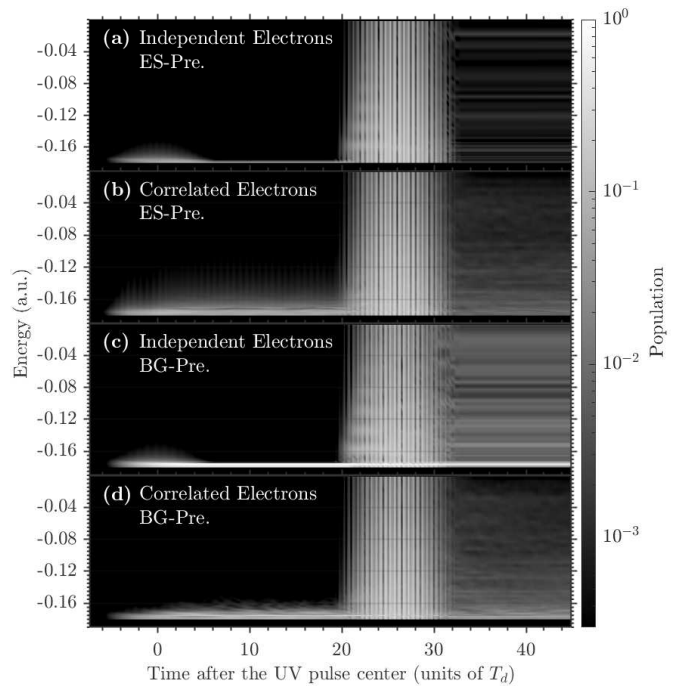}
\caption{Projections of the electronic wavepacket onto the initial bandstructure [using Eq.~\eqref{eq:proj}] during the simulation of the preexcited finite-size system ($N=80$) with parameters as of Fig.~\ref{fig:4}.
} \label{fig:5}
\end{figure}
To see whether the laser-induced energy shifts manifest directly in the evolution of electron populations, we have performed calculations similar to those for the analysis in Fig.~\ref{fig:2}(b). We apply Eq.~\eqref{eq:proj}, but for the preexcited systems of Fig.~\ref{fig:4}(a) and (c). We project the electronic wavepacket on the field-free KS orbitals at each timestep. The population distributed on the lower part of CB1 is given in Fig.~\ref{fig:5} as a function of time, for the laser pulse durations shown in Fig.~\ref{fig:3}(c), where the VUV preexcitation pump pulse acts on the system from $\sim -7.5$ to $\sim 7.5$ $T_d$ and the IR driving pulse acts on the system from $18.5$ to $33.5 T_d$. Since the basis and thus the bandstructure is static,  the time-dependent laser-induced  BG-energy shifts will be expressed through additional couplings, changing the electron populations. This can be observed by comparing Fig.~\ref{fig:5}(b) and (d) to (a) and (b) in the region of time $ < 18.5 T_d$ where the additional couplings allow the correlated calculations to populate a wider selection of the field-free states. In comparison, the static bandstructure of (a) and (c), exhibit a much more localized population of the states. Here the BG-preexcitation pulse [Fig.~\ref{fig:5}(c)] have more states within its spectral range, and thus allow a more effective population transfer, than the few states within the spectral width ($ < 0.0025$ a.u.) of the ES-preexcitation pulse [Fig.~\ref{fig:5}(a)]. If one considers the distribution of states at instants of time in-between the two pulses $7.5 T_d < t < 18.5 T_d$, a rather time-independent distribution of the independent-electron calculations of Fig.~\ref{fig:5} (a) and (c) is seen, whereas a time-dependence in the distribution of states is observed when including laser-induced correlations in (b) and (d). This difference suggests a $\tau$-dependence of the HHG spectra when including laser-induced correlations opposed to the independent-electron calculation, which is also observed in Ref.~\cite{jensen2020edgestateinduced}. We will return to this point in Sec.~\ref{Sec:3.2.3}. The driving pulse interacts with different carrier-doped systems after $18.5 T_d$. Here the intense driving pulse dominates the dynamics of the system within its duration, where the effects of carrier-doping only show as small changes in the frequency-region wherein the photo-carriers were initiated. After the time $33.5 T_d$ the driving field has ended, and the system end up with different population distributions of the non-recombined carriers depending on the initial preexcitation. This indicates that certain characteristics of the preexcitation step persist throughout the driving pulse. If one considers the higher CBs (not shown), a hauling-up-like increase of population is found during the IR-pulse. The basis of ground-state KS orbitals have been checked for completeness. In this connection we conclude that the preexcitation step results in a higher amount of ionization. 
As the population-transfer due to BG-preexcitation in Fig.~\ref{fig:5}(c) seems to be more distinct than that of the correlated calulation in Fig.~\ref{fig:5}(d), this may provide a reason for the large overestimation of the HHG-signal of the first plateau. 

\begin{figure}
\includegraphics[width=0.49\textwidth]{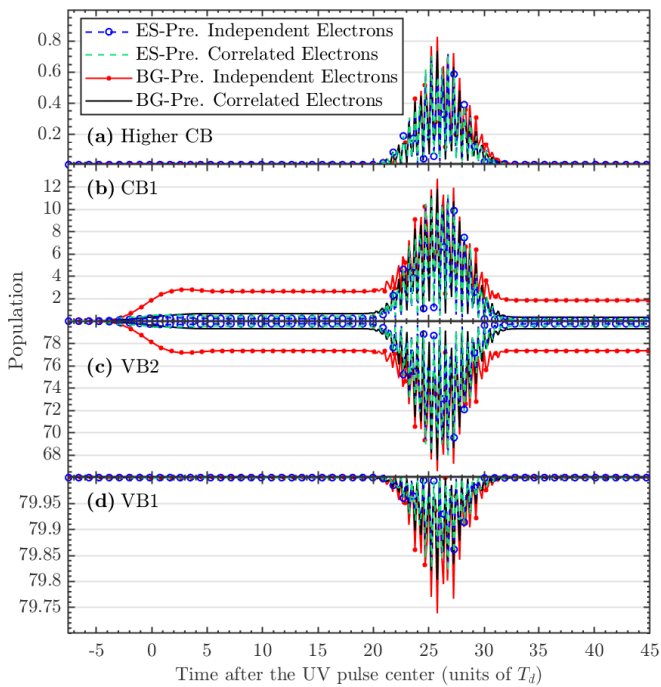}
\caption{Static bandstructure resolved electron populations during the simulation of the preexcited finite-size systems ($N=80$)  as given from the projections of Fig.~\ref{fig:5} and Eq.~\eqref{eq:proj_sum}. In (a) the population in all CBs higher than CB1 is shown. Parameters are defined as in Fig.~\ref{fig:4} and the notations VB1, VB2, CB1 are defined in Fig.~\ref{fig:1}.
} \label{fig:6}
\end{figure}

To quantify this speculation, similarly to Fig.~\ref{fig:2}(b), we have found the population of each VB and CB with Eq.~\eqref{eq:proj_sum} and provide the result in Fig.~\ref{fig:6}.
If we consider the excitation-ratio in Fig.~\ref{fig:6}, it is clear that CB1 and VB2 are responsible for most of the electron transitions. This is often used to justify simplified two-band models. As explained in Sec.~\ref{Sec:one color}, however, the lower-lying VBs can be important to include when considering laser-induced correlations. We see that the dominant source of transitions is the intense driving field, although, in the BG-preexcitation independent-electron calculation an immense population is already transferred from VB2 to CB1 during the preexcitation step. The degree of excitation that arises from the BG-preexcitation pulse is severely overestimated in the independent-electron approach. This overestimation shows that it is crucial to include laser-induced correlations if reporting on the controllability of carrier populations through resonant excitation.
The significant overestimation of the population transferred to CB1 does, however, not immediately transfer further onto the higher CBs of Fig.~\ref{fig:6}(a) if simply comparing to the correlated calculations. However, here it is important to note, that we would expect an increased population of the higher CBs in the correlated approach due to the hauling-up effect. Since the populations are of similar magnitude in the independent-electron and correlated calculations, we conclude that if a part of the overestimated population in the independent-electron calculation transfer further onto the higher bands, then this effect seems to match the magnitude of the hauling-up effect. It is clear that for the BG-resonant independent-electron calculation, only a part of the population of carriers from preexcitation is recombining, and a portion of it remains in the higher CBs even after the driving pulse. As concluded  in Sec.~\ref{Sec:one color} additional population in the CBs, does not strictly entail a higher HHG signal strength. If comparing the population at the beginning and end of the driving pulse we do, however, see that the overestimated preexcited population is expected to recombine emitting harmonics within the first plateau. Comparing with Fig.~\ref{fig:4} such an enhancement of the first plateau is exactly what is observed for the independent-electron case. We thus link the reduction of the yield in the HHG spectra due to laser-induced correlations for a BG preexcitation to stem from a reduction in the number of transferred carriers during the preexcitation step, as a result of dynamic change in the bandstructure. As the mechanism for this reduction is independent of system-size and can be explained through a momentum space bandstructure analysis, we denote it as a bulk-like laser-induced correlation effect. 

For both the finite-size ES-related laser-induced electron-correlation enhancement and the bulk-like laser-induced correlational reduction, the shared characteristics of laser-induced correlation is the interplay between the distribution of photo-carriers induced by the preexcitation pulse, and the interaction with the driving field. The laser-induced bulk-like correlational mechanisms for HHG are interpreted in two ways. Either as a widening of the distribution probed by the preexcitation pulse in a correlation-free basis, or as a time-dependency of a correlated bandstructure. No matter the interpretation, the laser-induced correlation effects will induce a time-dependent change of the distribution of preexcited carriers, and in the following section we will examine this time-dependency. 

\subsubsection{Laser-induced correlations effects in HHG as a function of time delay between preexcitation VUV pump and driving IR probe pulses\label{Sec:3.2.3}}
\begin{figure}
\includegraphics[width=0.49\textwidth]{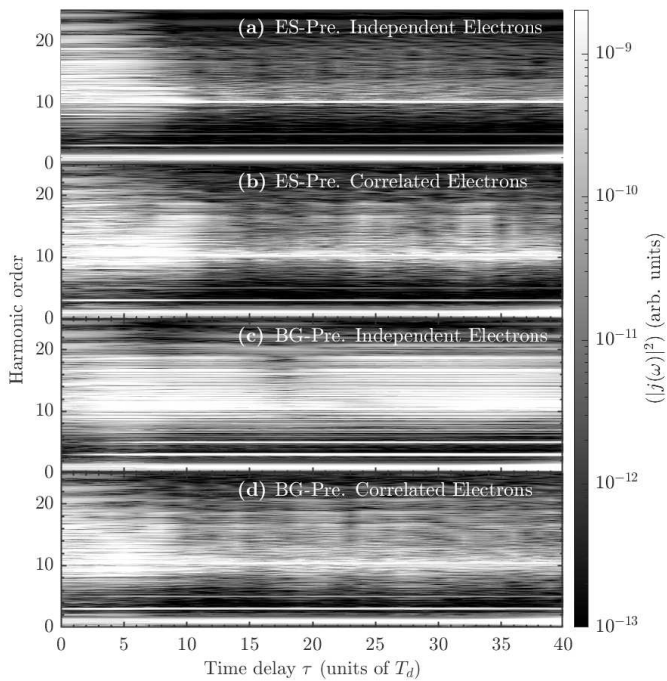}
\caption{(a) HHG spectra of a finite ($N=80$) system as a function of time-delay between preexcitation and driving pulse applying parameters as defined in Fig.~\ref{fig:4}. The signals have been scaled by $N^{-2}$.
} \label{fig:7}
\end{figure}
One could expect, in the limit of completely temporally overlapping pulses that there would be no significant laser-induced correlations since the phase of Eq.~\eqref{eq:timediff} would not have time to accumulate. Furthermore at $\tau = 0$ one could expect that the driving pulse might dominate the dynamics, since the driving IR pulse is two orders of magnitude more intense than the VUV, and the latter is in the perturbative regime. If the driving pulse completely dominates the dynamics, we return to the spectrum of the single-driving pulse studies of Sec.~\ref{Sec:one color} where much less significant effect of laser-induced correlations was seen. To approach this limit, and investigate the time-dependent dynamics of laser-induced correlations, a scan of varying time-delay is examined. The HHG-spectra are given in Fig.~\ref{fig:7}. Similarly as previously reported in Ref.~\cite{jensen2020edgestateinduced}, and as alluded to in the discussion of Fig.~\ref{fig:5}, the correlated approach show a more significant $\tau$-dependence throughout. The same pulses are applied as in the previous sections, even thought we note that applying a shorter driving pulse, as in Ref.~\cite{jensen2020edgestateinduced} will give access to a higher time resolution of variation in the HHG signal. Despite both pulses being of duration $\sim 15 T_d$ the correlated systems still exhibit a more detailed $\tau$-dependence which persist throughout large time-delays. The temporal structure of the spectra obtained by the uncorrelated, independent-electron approach seems more coarse, concluding, that the temporal structure of the HHG spectra contain information about the laser-induced 
correlational mechanisms. 
\begin{figure}
\includegraphics[width=0.49\textwidth]{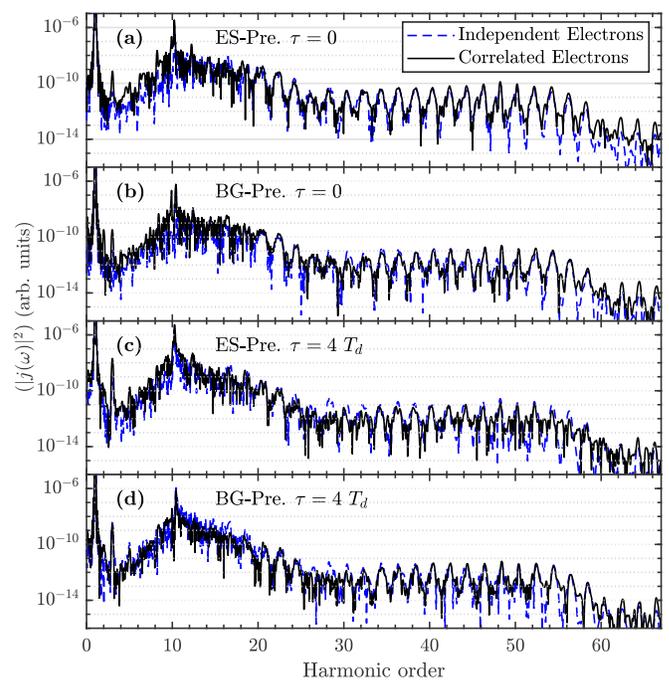}
\caption{HHG spectra for a finite-size system ($N=80$),  with independent-electrons and including laser-induced correlations for the ES-preexcitation (a) and (c) and for the BG-preexcitation (b) and (d) with $\tau = 0$ and $\tau = 4 T_d$. The parameters are identical to Fig. \ref{fig:7} of which, the the spectra are cross-sections at the given $\tau$. The signal has been scaled by $N^{-2}$. 
} \label{fig:8}
\end{figure}
In regions of large temporal overlap between the two pulses, laser-induced correlation effects are seen to be rather insignificant when considering the first plateaus of the HHG spectra. In this region laser-induced correlations only appear in higher plateaus as a hauling-up effect (not shown). The only exception to this trend is if considering intraband processes below the BG frequency, as these show laser-induced correlation effects at all $\tau$, see Fig. \ref{fig:8}. Thus we conclude that the independent-electron approximation breaks down at intraband processes in the VUV-pump--IR-probe HHG scenarios considered here. 
As we both observe laser-induced correlational enhancement and laser-induced correlational reduction of the spectra depending on the preexcitation frequency we can conclude that uncorrelated independent-electron calculations may  under- or overestimate the HHG signal when the pulses in a pump-probe setting are separated temporally. 

A qualitative comparison of the independent-electron calculations in Figs.~\ref{fig:7}(a) and (c) with the correlated approaches of Figs.~\ref{fig:7}(b) and (d), reveal that in order to find a significant difference in the uncorrelated and correlated spectra within the first plateau, a time delay of more than $4 T_d$ must be applied. This qualitatively set a scale for the laser-induced correlations. The delay of $4 T_d$ is depicted in Figs.~\ref{fig:8} (c) and (d), where, e.g., the $29.$ harmonic is reduced by an order of magnitude due to correlations. At this delay, the preexcitation pulse affects the system to induce laser-induced correlations in $\sim 25$ fs before the driving pulse starts and dominates the dynamics. Such a time delay fits well with the rough estimate of the time-scale for laser-induced correlations from Eq.~\eqref{eq:timediff}. 

\section{Conclusion}
In this work, we characterized laser-induced correlation effects in a linear chain model of a  BG material in two ways. Either in terms of nontrivial couplings in a static bandstructure picture resulting in a change in electron populations, or in terms of  time-dependent energy fluctuations of the states within the bandstructure. We distinguished between  two characteristic types of correlational effects. One is caused by the ESs. The presence of these states is a finite-size property. The population of the spatially localized ESs initiates density fluctuations. The associated dynamics requires a correlated description. The ES-induced correlation effect is reduced in relative importance as the size of the nanostructured linear chain increases. The other is a bulk-like laser-induced correlation effect, which can be explained by dynamical changes in the bandstructure that are independent of system size when the number of ions in the linear chain is sufficiently large ($N \gtrsim 60$), and which  induces a time-dependent change of the distribution of preexcited carriers. Laser-induced correlations were shown both to be able to strongly enhance and reduce the harmonic signal, depending on the applied laser pulses. The degree of excitation is a key parameter to determine the magnitude of laser-induced correlations. To this extent we indicated that  lower VBs, which are not producing a significant HHG signal, may become important to include if considering laser-induced correlations, as they are a source of nonexcited electrons, which will dampen the overall degree of excitation.

For single IR-pulse studies we observed an increased population of the higher-lying CB due to laser-induced correlations, which relates to the increased harmonic signal on the fourth and higher plateaus. This effect is typically not within the range of experimental detection. In a dynamic bandstructure analysis, the BG energy only changes in the range of a few percent during the interaction in the case of a single driving pulse, indicating a minor effect of laser-induced correlations in single-pulse schemes. If, however, a small change in the BG in a specific system would result in opening or closing of  multiphoton transitions, this general expectation could change. 

For VUV-pump--IR-probe pulse studies, we found that the VUV preexcitation pulse increases the degree of excitation and causes much stronger signatures of laser-induced correlations. We reported that laser-induced correlation effects are able to significantly enhance or reduce the HHG spectra across a wide range of harmonic frequencies. For intraband transitions below the BG energy, we find significant laser-induced correlation effect at all peak-to-peak delays between the VUV pump and the IR probe pulse. For interband transitions, producing harmonics above the BG-frequency, we found no laser-induced correlations for pulses that are applied simultaneously. If, on the other hand, the VUV pump arrives a few tens of femtoseconds before the IR probe, laser-induced correlations emerge. The laser-induced correlational effects exhibit a peak-to-peak delay-dependency and signatures hereof are seen in the  HHG spectra. Even though laser-induced correlations might increase the harmonic signal, we found  that the optimal enhancement is observed in the region of temporally overlapping pulses where the laser-induced correlational dynamics have not had time to build-up.
%
%
\begin{acknowledgements}
This work was supported by Danish Council for Independent Research (GrantNo.9040-00001B).
\end{acknowledgements}

%

\end{document}